# The Many Roads to Dementia: A Systems View of Alzheimer's Disease

Irina Kareva

Email: ikareva@asu.edu

**Abstract**

Alzheimer's disease is not the outcome of a single cause but the convergence of many. This review reframes dementia as a systemic failure, where amyloid plaques and tau tangles are not root causes but late-stage byproducts of the underlying metabolic collapse. We begin by tracing the historical merger of early- and late-onset Alzheimer's into a single disease category, a conceptual error that may have misdirected decades of research. We then synthesize evidence pointing to metabolic dysfunction - especially mitochondrial damage - as a more likely initiating event. Through this lens, we examine diverse contributing factors including type 2 diabetes, hyperglycemia-induced oxidative stress, infections and neuroinflammation. Finally, we assess current treatment limitations and argue that prevention, grounded in early metabolic and vascular interventions, holds the most promise for altering the course of this complex disease.

*Historical background: a tale of two dementias*

Alois Alzheimer first encountered Auguste Deter in 1901, when she was admitted to the Städtische Anstalt für Irre und Epileptische (Municipal Asylum for the Insane and Epileptics) in Frankfurt. She was 51 years old at the time of admission, and died in 1906, at the age of 55. She displayed severe cognitive impairments, including progressive memory loss, disorientation and confusion, language difficulties, hallucinations, and paranoia.

Alzheimer meticulously recorded her symptoms and, when she died in 1906, he collected and preserved tissue samples from her brain. These samples were later sectioned, stained, and mounted on slides for microscopic analysis in Emil Kraepelin's lab in Munich. Using newly developed staining techniques, he observed: amyloid plaques, which are extracellular deposits composed mainly of amyloid-β protein; neurofibrillary tangles (NFTs), which are twisted intracellular fibers composed of hyperphosphorylated tau protein; and cortical atrophy and neuronal loss.

He presented his findings in 1907 at the 37th Conference of South-West German Psychiatrists in Tübingen. His paper "Über eine eigenartige Erkrankung der Hirnrinde" ("On a peculiar disease of the cerebral cortex") was published in Allgemeine Zeitschrift für Psychiatrie und psychisch-gerichtliche Medizin (General Journal of Psychiatry and Psychological-Legal Medicine). In it, he proposed that this disease was distinct from "senile dementia," which was generally associated with aging (1).

While the symptoms of dementia had been described before, in Alzheimer's work the biological changes in the brain were observed for the first time under a microscope. This was made possible by advancements in silver staining techniques, which selectively highlight neurofibrillary tangles and amyloid plaques. Specifically, the Bielschowsky silver stain, developed by Max Bielschowsky in 1902-1903, was the critical method that enabled Alzheimer to clearly visualize plaques and neurofibrillary tangles. (Silver staining works by impregnating brain tissue with silver salts, which selectively bind to abnormal protein aggregates like amyloid plaques and neurofibrillary tangles. Unlike earlier methods, such as Nissl staining, developed by Franz Nissl, who worked in the same laboratory as Alzheimer, silver staining did not just highlight cell bodies but also revealed abnormal protein accumulations in finer detail. This allowed the structural pathology of Alzheimer's disease to be seen for the first time).

In the late 1990s and early 2000s, Auguste Deter's preserved slides were re-analyzed using modern immunohistochemical techniques. The 1997 study by R. Maurer et al. confirmed the presence of amyloid plaques and neurofibrillary tangles, consistent with Alzheimer's original observations (2). In subsequent years, it has been suggested that Deter carried a mutation in the PSEN1 (Presenilin-1) gene (3), which is associated with autosomal dominant early-onset Alzheimer's disease. However, a later re-evaluation (4) questioned this conclusion, citing degradation of the preserved tissue and methodological limitations that made definitive genetic analysis difficult. Therefore, whether Deter truly carried a PSEN1 mutation remains uncertain.

In 1907, the same year as Alzheimer's publication, Oskar Fischer, a Czech neuropathologist, described 16 cases of senile dementia (5) at the German Charles-Ferdinand University in Prague (the institution was later renamed Charles University in Prague after World War I and still exists today). Fischer studied elderly patients (aged over 65, unlike Alzheimer's much younger patient). He described plaques in aging brains and noted dystrophic neurites, which are damaged and swollen neuronal processes surrounding the plaques. Fischer suggested that plaques were linked to cognitive decline in aging, though he did not propose a definitive causal mechanism. Unlike Alzheimer, Fischer did not identify neurofibrillary tangles, but his description of the plaques was more systematic (6).

Emil Kraepelin, in whose lab Alois Alzheimer worked at the time, was a highly influential psychiatrist, who played a major role in popularizing Alzheimer's discovery. In fact, it was he who in his 1910 textbook coined the term Alzheimer's disease, referring specifically to Auguste Deter's early-onset case. Fischer, despite his meticulous research and a larger sample size, did not receive similar recognition, likely due to him working in a much less prestigious institution. His Jewish background likely also played a role in the historical marginalization of his scientific contributions, especially in light of subsequent historical events.

Originally, the term Alzheimer's disease referred only to early-onset cases (Auguste Deter's). Later, as more cases of "senile dementia" were examined, amyloid plaques were found in both young and old patients. Over time, as more elderly patients were found to have plaques and tangles in their brains, researchers merged early-onset and late-onset cases under the single diagnosis of Alzheimer's disease (7).

This classification, however, appears to have conflated two different conditions. Early onset Alzheimer's disease is largely genetic: it is associated with mutations in PSEN1 (8), PSEN2 (9) and APP (10,11) genes, which, while accounting only for under 5% of all Alzheimer's patients, are highly penetrant (i.e., an individual with these mutations is very likely to develop early onset Alzheimer's disease). In contrast, late-onset Alzheimer's disease, which accounts for ~95% of cases, is influenced by a combination of genetic, environmental, and lifestyle factors. While specific mutations do not cause late onset Alzheimer's disease in the same way as early onset, one genetic factor—the ApoE4 allele—is known to increase risk. Unlike PSEN1, PSEN2, and APP mutations, ApoE4 is not fully penetrant, meaning that not everyone who carries it develops the disease, but it significantly raises susceptibility.

*History and function of ApoE*

Apolipoprotein E (ApoE) is a critical protein involved in lipid metabolism and neural function (12). ApoE is primarily synthesized in the liver, where it plays a significant role in transporting lipids, including cholesterol and triglycerides, through the bloodstream to various tissues (13). Within the brain, ApoE is predominantly produced by astrocytes and, to a lesser extent, by microglia (14). Neurons generally do not produce ApoE but express receptors to take up ApoE-bound lipids (13). While peripheral ApoE plays a role in systemic lipid transport, it does not significantly cross the BBB under normal conditions. Therefore, ApoE required in the CNS is synthesized locally by glial cells and astrocytes.

ApoE proteins bind to receptors like LDL receptor (LDLR), LDL receptor-related protein 1 (LRP1), VLDL receptor, and ApoER2 (15). There exist three isoforms of the ApoE protein: 2,3 and 4, which structurally differ in amino acid substitutions in positions 112 and 158 (16). They differ in their affinity for their targets (i.e., how tightly they bind to them), with ApoE4 having the tightest binding, and ApoE2 and lightest. ApoE4 carriers appear to be more susceptible to late onset Alzheimer's disease, while ApoE2 appears to be protective, relative to the most common "neutral" variant ApoE3.

ApoE proteins carry a cargo of cholesterol and lipids. Normally (for the ApoE3 variant), when an ApoE protein binds to its receptor, the protein-receptor complex gets internalized via endocytosis. Inside the cell endosome, lipoproteins are broken down into free cholesterol and phospholipids, which are essential for synapse function, myelin repair, neuronal membrane integrity, and lipid raft formation. The ApoE protein, after releasing its cargo, detaches and is either recycled or degraded; the receptor itself if recycled back to the membrane to continue lipid uptake (17).

Because the affinity of ApoE4 to its targets is higher, it binds more tightly to the receptor. When this protein-receptor complex gets internalized, instead of recycling, the receptor is trafficked to lysosomes for degradation. Therefore, tighter binding over time leads to decrease in receptor expression (since receptors that have interacted with ApoE4 do not get recycled but are degraded), which paradoxically long-term leads to lipid deprivation and decreases the cells' ability to repair themselves. Since cholesterol is critical for synapse function and myelin repair, when neurons receive insufficient cholesterol due to ApoE4 activity, it stresses the cells and activates inflammatory pathways. Lipid droplets accumulate in glia due to impaired lipid trafficking, increased oxidation of lipids, and neuroinflammatory activation of microglia, triggering release of inflammatory cytokines like TNF-alpha, IL-1β and IL-6, propagating the cycle of neuroinflammation.

In contrast, ApoE2 binds less tightly to lipoprotein receptors, leading to slower cholesterol uptake. However, it does not interfere with receptor recycling, ensuring that neurons still receive adequate cholesterol without triggering oxidative stress or inflammation (18). The ApoE2 variant may have emerged in populations with lower infectious disease burden, where excessive inflammation was less advantageous. These mechanisms and differences between the variants are summarized in Figure A1.

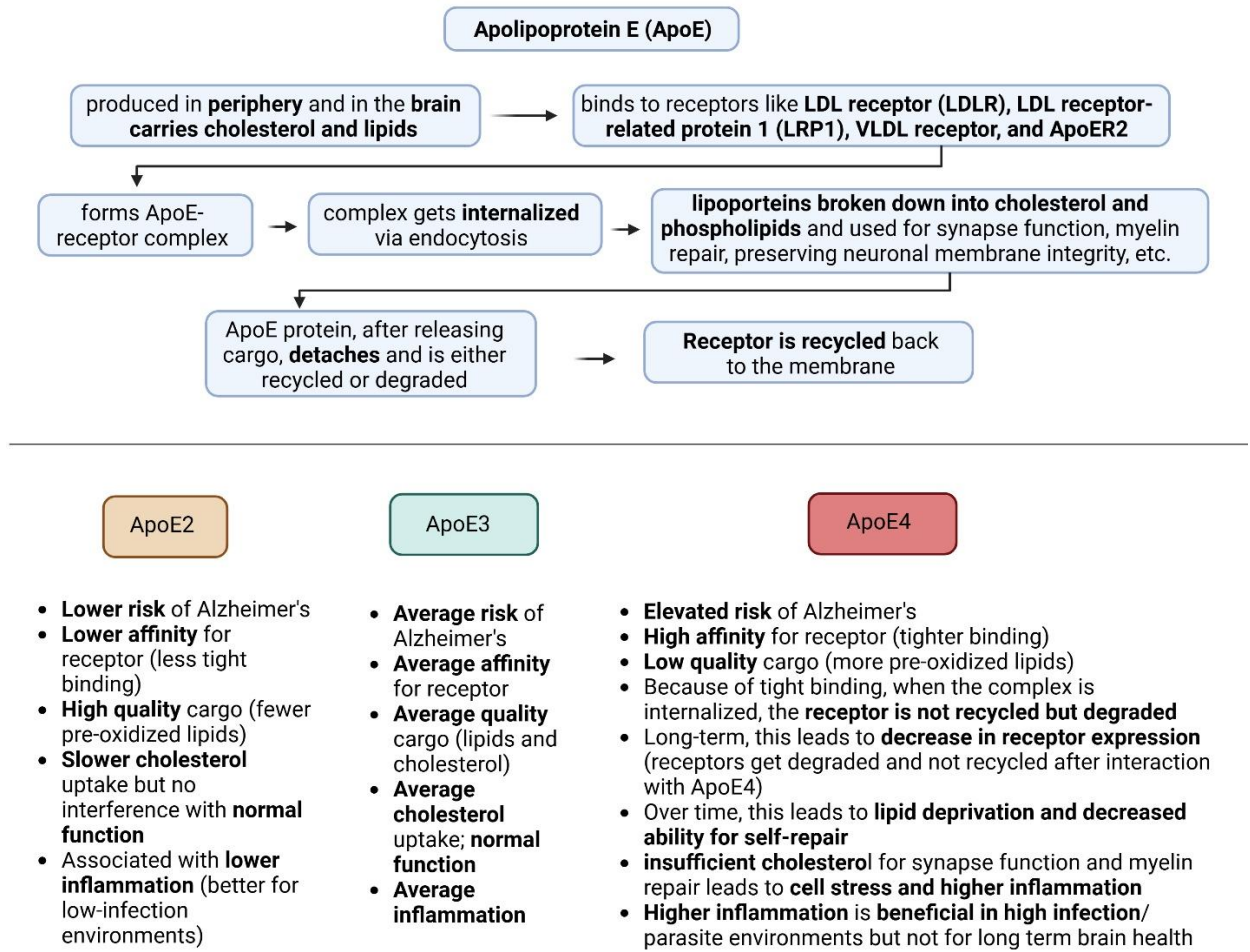

**Figure 1**. Summary of the normal function of the ApoE protein, and the differences between the ApoE isoforms.

Interestingly, ApoE4 used to be a dominant variant in the population until approximately 10000 years ago. Today, ApoE4 is found at higher frequencies in tropical latitudes and in some hunter-gatherer and forager-horticulturalist populations. For example, research on the Tsimane, an Amazonian forager-horticulturalist group, suggests that ApoE4 may be protective in environments with high pathogen loads (19). Specifically, ApoE4 is associated with a reduced parasite burden and lower eosinophil counts, a marker of parasitic infection. Furthermore, in (19), the authors suggest that ApoE4 carriers show improved clearance of certain infections, such as hepatitis C and cryptosporidium. Of course, while ApoE4 may have been adaptive for survival in parasite-heavy environments, its effects in modern, low-infection societies become maladaptive. Interestingly, ApoE4 carriers in high-pathogen environments (e.g., the Tsimane people) show better cognitive function than non-carriers, suggesting its deleterious effects are context-dependent.

*Different hypotheses about the causes of late onset Alzheimer's disease*

For decades, the dominant hypothesis has been that amyloid β plaques are the primary cause of Alzheimer's disease. As mentioned above, this view was driven by the identification of amyloid β plaques in the brains of Alzheimer's patients through post-mortem histology using silver-staining techniques.

Amyloid-β (Aβ) is a fragment of amyloid precursor protein (APP), a transmembrane protein involved in neuronal growth and repair. It exists in monomeric, oligomeric, and fibrillar forms. Monomers are single Aβ molecules, oligomers are soluble aggregates of a few Aβ molecules, and fibrils are insoluble, larger aggregates that accumulate to form amyloid plaques. Some evidence suggests that oligomeric amyloid-β, rather than fibrillar plaques, may be the most neurotoxic species (20).

APP is cleaved by two enzymes, β-secretase (BACE1) and gamma-secretase, to produce amyloid-β peptides of varying lengths (ranging from Aβ38 to Aβ43). The number (e.g., Aβ40, Aβ42, Aβ56) refers to the number of amino acids in the peptide. Among these, Aβ42 is the most aggregation-prone form, making it the primary component of plaques in Alzheimer's disease. By contrast, Aβ40 is more soluble and less likely to form toxic aggregates. CSF amyloid-β 42 levels are normally ~500 pg/mL, with levels below ~400 pg/mL often considered pathological. Counterintuitively, CSF amyloid-β 42 levels decrease in AD, as the peptide is sequestered into plaques in the brain, reducing its presence in cerebrospinal fluid (21). This biomarker is now widely used in the clinical diagnosis of Alzheimer's disease.

Historically, amyloid β plaques were detectable only in post-mortem brain tissue. However, since the 1990s, positron emission tomography imaging with Pittsburgh Compound B (PiB-PET) has allowed amyloid to be detected in living patients. PiB is a radiolabeled compound that binds specifically to fibrillar amyloid-β deposits in the brain, allowing visualization through PET scans (22,23). However, PiB-PET does not detect soluble oligomeric amyloid-β.

Since the early 2000s, it has become possible to detect amyloid β in cerebrospinal fluid (CSF) and blood. Recent advancements have also led to blood tests capable of predicting the presence of β-amyloid plaques in the brain with high accuracy (24–27), but these measurements have limited predictive power for AD progression. A major challenge to the amyloid hypothesis has been the observation that approximately 30% of elderly people without dementia have extensive amyloid plaques but no symptoms of cognitive decline (28). In fact, their brains can look indistinguishable from those of Alzheimer's patients in terms of plaque deposition. Interestingly, while most Alzheimer's patients have amyloid plaques, approximately 10–15% of clinically diagnosed AD patients exhibit little to no amyloid burden upon autopsy, a condition referred to as plaque-negative AD (29,30).

This raises the question of whether tangles and plagues in the brains of people with dementia are byproducts of pathology rather than their cause. If plaques were the cause, we would expect anti-amyloid drugs to reverse or stop Alzheimer's progression. However, this has not been the case. In fact, recently approved anti-amyloid drugs, such as aducanumab and lecanemab, reduce plaques but show only marginal clinical benefits (31,32). For instance, in EMERGE and ENGAGE trials for aducanumab, 70% of the plaques were cleared, yet this did not affect cognitive decline (33). Similarly, in CLARITY-AD trial for lecanemab, reduced amyloid plaques by approximately 70% but decline was slowed only by about 27%, which translates to a delay in progression of only approximately 4.5 months over an 18-month period, a modest benefit at best (31). It has been hypothesized that perhaps the interventions were taken too late in the disease process, although anti-amyloid trials in genetic early onset Alzheimer's disease also failed to significantly prevent disease progression (34). Some recent results suggest that while partial and short term removal of amyloid deposits does not show significant efficacy, long term administration of gantenerumab (another anti-amyloid drug) may provide some clinical benefit but it remains to be investigated (35).

Interestingly, rather than being a purely toxic agent, amyloid-β may actually serve important physiological functions, including protecting the BBB from microbial invasion by acting as an anti-microbial peptide (36). It binds microbial surfaces, aggregated into fibrils that trap pathogens and forms a physical barrier at the BBB. One of the most troubling side effects of the anti-amyloid drugs are brain bleeds (amyloid-related imaging abnormalities, ARIA), and it can be hypothesized that disrupting amyloid-β's protective barrier function at the BBB may contribute to vascular fragility.

Amyloid-β is involved in synaptic regulation by modulating N-Methyl-D-Aspartate (NMDA) receptor activity. NMDA receptors are critical for synaptic plasticity and memory function (37). At physiological levels, amyloid-β can enhance synaptic plasticity and memory; however, elevated levels lead to synaptic dysfunction and cognitive deficits.

Some studies have also demonstrated that sleep deprivation increases amyloid-β levels in the brain interstitial fluid, as measured by microdialysis techniques in animal models (38). This occurs because sleep enhances glymphatic clearance of amyloid-β, while wakefulness leads to progressive accumulation. In humans, CSF studies have similarly shown that amyloid-β concentrations fluctuate with the sleep-wake cycle (39–41), reinforcing the hypothesis that impaired clearance mechanisms during sleep may contribute to Alzheimer's pathogenesis.

Notably, in 2022, an investigation published in Science (Schrag et al., 2022) exposed potential data manipulation in a 2006 paper that strongly promoted the amyloid hypothesis. The now retracted paper in question, published in Nature (42), claimed that a specific amyloid-β oligomer (Aβ*56) was responsible for memory deficits in mice. However, image analysis suggested that key findings were based on manipulated Western blots, leading to concerns that this study had misled the field for nearly two decades. While this does not disprove the amyloid hypothesis outright, it raises concerns that funding and research priorities may have been overly focused on amyloid at the expense of other theories.

Rather than focusing only on plaque formation, an alternative view suggests that the problem lies in the inability to clear amyloid-β effectively. It has been hypothesized that the brain's waste-clearing system (the glymphatic system), particularly during sleep, may be impaired in Alzheimer's (43). It's also been suggested that, while microglia (immune cells in the brain) normally clear amyloid-β, in aging and disease they may become dysfunctional or hyperactive, leading to chronic inflammation (44), which in turn can propagate the vicious cycle of damage that will be discussed in more detail later. Reduced CSF drainage has also been associated with a higher amyloid burden (45). This could explain why drugs that simply remove amyloid-β plaques are ineffective, if the real issue is systemic failure of brain maintenance mechanisms.

*The tau hypothesis: the neurofibrillary tangles*

Recall that the key observations made by Alois Alzheimer on the slides of Auguste Deter's tissues were amyloid plaques, regions of neuronal loss, and neurofibrillary tangles (NFTs).

Tau is a microtubule-associated protein that stabilizes neuronal cytoskeletal structures and facilitates axonal transport by binding to microtubules. Under normal conditions, tau undergoes regulated phosphorylation at specific sites, allowing for dynamic attachment and detachment from microtubules to facilitate cellular transport. Normally, tau phosphorylation is tightly regulated by kinases (e.g., GSK-3β, CDK5) and phosphatases (e.g., PP2A). However, in Alzheimer's disease, increased kinase activity and reduced phosphatase activity lead to hyperphosphorylation, causing tau to detach from microtubules and aggregate into NFTs. This process disrupts the intracellular transport of organelles (e.g., mitochondria), nutrients, and neurotransmitters (46). It also contributes to synaptic dysfunction, leading to neurodegeneration. Unlike amyloid-β plaques, which accumulate extracellularly, NFTs form inside neurons, directly interfering with cellular function. Eventually, affected neurons undergo apoptosis (programmed cell death), exacerbating cognitive decline. Some studies suggest that NFT burden correlates more closely with cognitive decline than amyloid deposition (47).

Currently, there are no widely approved tau-targeting drugs. Existing tau antibodies, such as gosuranemab and semorinemab, were designed to clear extracellular tau but have shown limited cognitive benefits in clinical trials (48,49). Tau kinase inhibitors, such as tideglusib, aim to block hyperphosphorylation, but concerns over off-target effects and toxicity have limited their development (50,51). Another approach involves microtubule stabilizers, such as Epothilone D, which help reinforce the cytoskeleton by preventing tau-induced disassembly of microtubules, though they have yet to demonstrate clear clinical success (52). These failures suggest that tau pathology may be a downstream effect of broader neurodegenerative processes, rather than the primary disease driver.

Like amyloid-β, tau tangles may be a later-stage effect of a more fundamental disease process. Some researchers argue that tau pathology spreads primarily after neuronal stress and synaptic damage have begun, rather than initiating the disease itself. For instance, (53) demonstrated that tau aggregation correlates with pre-existing synaptic loss, suggesting that tau pathology follows neuronal injury. Similarly, (54) proposed that tau spreads in a prion-like manner, where misfolded tau induces normal tau to misfold and aggregate. This pathological spread occurs through synaptically connected neurons, contributing to disease progression (55). However, tau misfolding likely requires an initial trigger, such as metabolic dysfunction, oxidative stress, or chronic neuroinflammation (56). These findings challenge the idea that tau tangles are the primary cause of AD, suggesting they may instead be a downstream consequence of broader neurodegenerative processes and potentially part of a self-perpetuating feedback loop.

*The energy crisis hypothesis: is Alzheimer's a metabolic disorder?*

A competing hypothesis, championed by researchers such as Jack de la Torre in the 1990s and 2000s (57,58), suggests that Alzheimer's disease, along with other neurodegenerative disorders, is primarily driven by chronic cerebral hypoperfusion and metabolic dysfunction, rather than amyloid accumulation. However, this view was largely overshadowed by the amyloid cascade hypothesis, which gained dominance due to the discovery of APP, PSEN1, and PSEN2 mutations in early-onset AD (1991–1996) and the subsequent prioritization of amyloid-focused research funding. The now-retracted Lesné et al. (2006) study further reinforced the amyloid-centric view, delaying broader consideration of alternative hypotheses such as the metabolic model.

The human brain contains ~86 billion neurons and an estimated 100 trillion synapses (59,60). These neurons require enormous amounts of energy to maintain synaptic function and cellular homeostasis. The brain consumes ~20% of the body's total energy supply, despite being only 2% of body weight. Cerebral blood flow (CBF) naturally declines with age, decreasing by approximately 0.5% per year, leading to a ~20% reduction from young adulthood (~22 years) to midlife (~62 years) (61). While this decline is normal, additional reductions – whether due to vascular disease, metabolic dysfunction, or neuroinflammation – can impair neurovascular coupling, the mechanism that ensures active brain regions receive sufficient oxygen and glucose. Chronic disruption of this process has been implicated in the development of Alzheimer's disease and other dementias.

Mitochondrial dysfunction and glucose metabolism abnormalities are well-documented in Alzheimer's patients (62–64). Evidence from PET imaging studies shows that glucose hypometabolism precedes amyloid deposition in Alzheimer's disease, suggesting metabolic failure is an early event. In (65), the authors demonstrated that reductions in glucose uptake in the posterior cingulate cortex and temporoparietal regions were present before significant amyloid plaque formation. This supports the hypothesis that metabolic dysfunction, rather than amyloid accumulation, may be the initiating factor in late-onset Alzheimer's disease.

Perhaps one of the most compelling experiments supporting the metabolic hypothesis was conducted by (66), where the authors partially inhibited cytochrome c oxidase (COX, Complex IV of the electron transport chain) using sodium azide, a weaker analog of cyanide. COX is the final enzyme in the mitochondrial electron transport chain, responsible for transferring electrons from cytochrome c to oxygen, a critical step in ATP production. When COX activity is impaired, neurons experience an energy deficit, leading to synaptic dysfunction, oxidative stress, and ultimately cell death. A 35% reduction in COX activity resulted in memory deficits resembling AD, but no amyloid plaque accumulation, suggesting that neuronal energy failure precedes amyloid pathology. This aligns with PET findings showing glucose hypometabolism in AD before amyloid or tau deposition (65,67).

Interestingly, emerging evidence suggests that metabolic dysfunction is a common factor in multiple neurodegenerative disorders, although the affected brain regions differ. Vascular dementia results from chronic cerebral hypoperfusion, particularly affecting white matter and subcortical structures (57,68). Alzheimer's disease primarily affects the hippocampus and temporoparietal cortex, with glucose hypometabolism appearing before amyloid plaques (65). Lewy body dementia (LBD) is associated with occipital cortex hypometabolism (69), while Parkinson's disease (PD) exhibits mitochondrial dysfunction and reduced glucose uptake in the basal ganglia (70). Despite their distinct presentations, these diseases share underlying metabolic impairments, suggesting that neurodegeneration may arise from regional energy deficits rather than a single toxic protein.

Alzheimer's disease has also been referred to as "Type 3 Diabetes" (71,72) due to the presence of insulin resistance in the brain. Studies have shown that AD brains exhibit reduced insulin receptor expression, impaired glucose uptake, and mitochondrial dysfunction, drawing parallels to Type 2 Diabetes (73). Indeed, there exists significant co-morbidity between T2D and Alzheimer's disease, which we will discuss in detail next.

*Type 2 Diabetes and The Metabolic Hypotheses*

Paradoxically, high systemic blood glucose may result in low availability of functional glucose in the brain.

The brain is protected by a thick membrane called the blood brain barrier (BBB), which strictly regulates access to the brain of any systemic toxins and chemicals and which ensures that there is always sufficient glucose available for brain cells. One way that glucose homeostasis is regulated is through insulin-independent GLUT1 transporters, i.e., they do not require insulin to transport glucose from systemic circulation into the brain extracellular space (74). GLUT1 transporters, which line the BBB and have a high affinity for glucose, help prioritize glucose delivery to the brain, especially when systemic glucose is low. However, chronic hyperglycemia can lead to impaired GLUT1 function and distribution, reducing effective glucose transport into the brain (75). As such, even when systemic blood glucose may be high, baseline glucose levels in the brain become less tightly regulated, and may become lower compared to healthy individuals, or even oscillate, with periods of excess glucose influx alternating with periods of deprivation (76). If systemic glucose levels fluctuate or remain very high for an extended period of time, as in individuals with T2D, this regulatory system may start to break down, leading to increased permeability of the BBB (also known as "leaky" BBB).

Chronic systemic hyperglycemia can lead to both oxidative stress and formation of advanced glycation end products (AGEs), which form when proteins, lipids, or nucleic acids react with sugars. AGEs damage endothelial cells of the BBB, increase its permeability, and promote inflammatory cytokine release, allowing toxins and pro-inflammatory molecules to enter the brain while disrupting glucose transport regulation (77,78). However, one of the most damaging consequences of hyperglycemia is loss of small blood vessels (capillary rarefaction), which leads to reduced oxygen and nutrient delivery to neurons (79,80). This means that even if glucose is present in the blood and the extracellular space in the brain, it may not be able to effectively reach neurons due to poor cerebral perfusion. Interestingly, PET scans using fluorodeoxyglucose (FDG-PET) show reduced glucose metabolism in neurodegenerative diseases. For instance, areas like the hippocampus and posterior cingulate cortex are particularly affected in Alzheimer's disease, with hypometabolism detected before significant amyloid deposition (81,82). Hyperglycemia in the brain might also affect astrocytes, cells that support neural function by converting glucose into lactate and transferring it to neurons via the astrocyte-neuron lactate shuttle (ANLS), which neurons then use for energy (83,84). Most neuronal energy is obtained through insulin-independent GLUT3-mediated glucose uptake, which fuels the TCA cycle (85). However, astrocytes, which rely on both insulin-independent GLUT1 and insulin-dependent GLUT4 transporters, can additionally augment neural energy supply with lactate, particularly under high activity, metabolic stress, or injury (86). Chronic hyperglycemia can lead to astrogliosis, an abnormal increase in the number, shape, and function of astrocytes in response to injury, inflammation, or metabolic stress (87). Astrogliosis disrupts normal astrocyte functions, leading to excess production of inflammatory cytokines (such as IL-6 and TNF-alpha), chronic neuroinflammation, and impaired metabolic support for neurons (86).

Importantly, astrocytes also play a critical role in glutamate clearance, preventing excessive excitatory signaling (84,88). When astrocytes become dysfunctional, their ability to remove excess glutamate from the synaptic space is compromised, leading to glutamate excitotoxicity—a process in which neurons are damaged or killed by excessive activation of glutamate receptors (89). High glutamate levels overstimulate NMDA receptors, resulting in excessive calcium influx into neurons. This calcium overload triggers mitochondrial stress, excessive prodiction of reactive oxygen species (ROS), and ultimately neuronal death, further exacerbating neurodegeneration (90). The combination of energy failure, inflammation, and glutamate excitotoxicity creates a vicious cycle, compounding neuronal damage and accelerating the progression of neurodegenerative diseases.

Finally, glucose oscillations in the brain, which can start occurring due to increased leakiness of the BBB, can lead to mitochondrial dysfunction (91,92). Unlike muscle or liver cells, neurons do not store glucose when glucose is in excess, and do not revert to glycolysis when glucose is low but instead depend on a steady nutrient supply. As a result, periods of excessive glucose flux (hyperglycemia) can lead to mitochondrial stress, while periods of nutrient deficiency due to cerebral hypoperfusion or astrocyte dysfunction lead to neuronal starvation (93,94). This mismatch between glucose supply and demand can lead to mitochondrial failure, which is central to neuronal energy production since, as noted above, neural cells cannot switch to glycolysis and instead depend on the TCA cycle for energy.

Specifically, leaky BBB can flood neurons with high glucose, overloading the mitochondria and leading to excessive electron transport chain (ETC) activity, which in turn produces excessive reactive oxygen species (ROS) (94). High levels of ROS, together with AGEs, damage mitochondrial proteins, including those of the ETC, impairing ATP production and causing mutations in mitochondrial DNA, which lacks robust repair mechanisms, leading to dysfunction accumulation (77). ROS also damages lipid membranes, including cardiolipin, a phospholipid crucial for ETC stability, which predisposes neurons to apoptosis (95). Additionally, ROS from dysfunctional mitochondria can further damage endothelial cells in the BBB, increasing its leakiness and therefore propagating a cycle of glucose fluctuation, inflammation, and oxidative stress, reinforcing the damage (96,97). Finally, leaky mitochondria also release pro-inflammatory signals. Recall that mitochondria originated from an ancient bacterial ancestor that entered into a symbiotic relationship with eukaryotic cells approximately 1.5 billion years ago (98). Because of its bacterial origin, mitochondrial DNA and proteins resemble bacterial components, triggering immune responses when released from damaged cells (99,100). When neurons are damaged, mitochondrial damage-associated molecular patterns (mtDAMPs) are released, activating microglia, the immune cells of the brain (101,102). Microglia recognize mtDAMPs as foreign molecules, leading to inflammasome activation, with increased TNF-alpha, IL-6, and IL-1β levels. This response amplifies oxidative stress, increases amyloid-β and tau accumulation, and further damages mitochondria, creating a vicious cycle (102).

As was mentioned above, emerging evidence suggests that both amyloid-β and tau have normal evolutionary protective functions, particularly in sealing leaks in the blood-brain barrier (BBB) during early neurodegeneration (36). Amyloid-β interacts with endothelial cells and is deposited at sites of BBB damage, potentially reinforcing weakened areas (103). Meanwhile, tau's primary function is to stabilize microtubules, the internal scaffolding system of neurons, which facilitates transport of mitochondria, nutrients, and cargo (104,105). Under normal conditions, tau undergoes transient phosphorylation to regulate its function, allowing microtubules to dynamically remodel. However, chronic inflammation, ROS, and metabolic stress trigger tau hyperphosphorylation, causing it to detach from microtubules. Once detached, tau misfolds and aggregates into toxic oligomers, disrupting axonal transport, particularly mitochondrial transport to synapses, further worsening neuronal energy failure (106,107). It has been hypothesized that tau may have evolved as a stress-response protein, temporarily detaching from microtubules to allow for rapid cytoskeletal restructuring under injury conditions (108). However, in disease states, persistent hyperphosphorylation and aggregation prevent tau from returning to its normal function, leading to axonal degeneration and worsening neurodegeneration.

Unfortunately, as these (potentially initially protective) proteins accumulate, they may start causing damage, with amyloid β binding mitochondrial membranes, further impairing the electron transport chain (complexes I and IV), increasing ROS production, mitochondrial permeability and leading to cytochrome c release and apoptosis. Amyloid β and tau together disrupt the delicate balance between mitochondrial fusion (which promotes repair and ATP efficiency) and fission (which removes damaged mitochondria). This leads to excessive fragmentation, impairing mitochondrial energy production and synaptic function. Without proper transport, mitochondria fail to reach synapses, starving them of ATP and leading to synaptic loss. As mitochondrial damage accumulates, neurons fail to produce enough ATP, which leads to weakened synaptic connections due to impaired energy supply. This ultimately can lead to neural death, accelerating dementia progression.

The key features of these complex feedback loops are summarized in Figure 2.

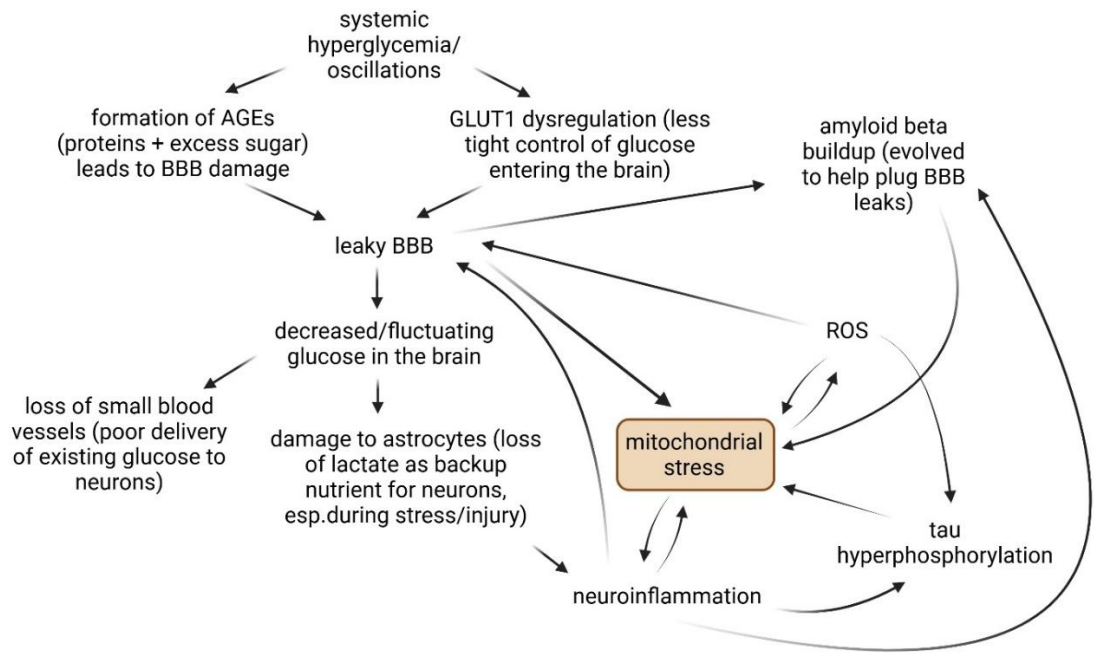

*Figure 2.* Key features of the complex feedback loops leading to neurodegeneration.

*The many roads to dementia*

However, T2D is not the only mechanism, whereby these processes can be triggered and amplified. For instance, carriers of the ApoE4 allele might have higher baseline inflammation, coupled with reduced cholesterol and lipids available for neural cells to sustain and repair themselves. ApoE4 can also promote amyloid-β buildup & tau hyperphosphorylation, which further damage mitochondria, leading to a vicious cycle and making the brain more vulnerable to neurodegeneration through amplified mitochondrial stress & inflammation.

Other factors that can initiate and augment mitochondrial damage include head trauma, polypharmacy, and infections. For instance, traumatic brain injury (TBI) can cause excessive calcium influx into neurons, triggering the opening of the mitochondrial permeability transition pore (mPTP), which leads to ATP depletion, mitochondrial swelling, and ultimately cell death (109,110). Furthermore, tau pathology accelerates in TBI patients (e.g., chronic traumatic encephalopathy in football players (111,112)). Strokes and ischemic injuries, which block glucose and oxygen delivery to parts of the brain, can result in energy crisis through mitochondrial failure due to lack of oxygen. Additionally, reperfusion injury occurs when blood flow returns to previously oxygen-deprived tissue (such as after a stroke or cardiac arrest). This sudden return of oxygen overloads mitochondria, leading to an explosion of ROS production due to excess electron transport chain activity. This oxidative burst damages mitochondrial membranes, leading to cell death and further inflammation (113,114).

Polypharmacy – multiple medications use – could additionally cause symptoms that may resemble dementia (115). For instance, statins can reduce CoQ10, which is critical for ETC function (116); benzodiazepines might reduce mitochondrial ATP production (117); some anticholinergic drugs, used for allergy, depression and bladder control, might also interfere with mitochondrial function (118).

Finally, infections are emerging as potential contributors to neurodegeneration. Forn instance, chronic periodontal infections have been linked to an elevated risk of dementia (119–121). The bacterium *Porphyromonas gingivalis*, commonly implicated in gum disease, has been detected in the brains of Alzheimer's patients; its presence is associated with increased amyloid β accumulation (122). The proposed mechanism involves systemic inflammation and direct bacterial invasion leading to neuroinflammation and neuronal damage.

HSV-1, known for causing cold sores, has been found in the brains of many elderly individuals, particularly those carrying the ApoE4 allele (123,124). Reactivation of latent HSV-1 in the brain can lead to neuroinflammation, contributing to amyloid plaque formation and tau pathology, thereby increasing the risk of Alzheimer's disease.

Some chronic inflammatory disorders have also been associated with increased risk of dementia. For instance, Lyme disease, caused by *Borrelia burgdorferi*, can lead to chronic neuroinflammation if not adequately treated (125,126). Persistent infection may result in cognitive impairments and has been associated with an increased risk of dementia.

Post-acute sequelae of SARS-CoV-2 infection, commonly referred to as long COVID, have also been linked to persistent cognitive deficits (127). The underlying mechanisms are thought to involve chronic inflammation, direct viral invasion of neural tissues, and disruption of the blood-brain barrier, all of which can precipitate or exacerbate neurodegenerative processes (128,129).

Severe systemic infections, such as pneumonia, urinary tract infections, and sepsis, have been associated with an increased risk of subsequent dementia (130,131). The systemic inflammatory response elicited by these infections can lead to neuroinflammation, thereby contributing to cognitive decline (132).

The connection between infections and neuronal energy deficits can be understood through several interrelated mechanisms. Infections trigger the release of pro-inflammatory cytokines as part of the body's immune response, which can cross the blood-brain barrier, leading to neuroinflammation. Chronic neuroinflammation can impair mitochondrial function within neurons, reducing ATP production and precipitating an energy crisis that compromises neuronal survival and function. Additionally, certain pathogens have the ability to directly invade neural tissues, like the HSV-1 virus mentioned above. Finally, activated immune cells exhibit increased glycolytic activity to meet their energy demands during an immune response (133). This heightened glucose consumption could potentially reduce the availability of glucose for neurons, which rely on oxidative phosphorylation for energy production. The resultant energy deficit in neurons can impair their function and viability, potentially leading to cognitive impairments, if the infection and inflammation continues past the acute and into the chronic phase. This is summarized in Figure 7.

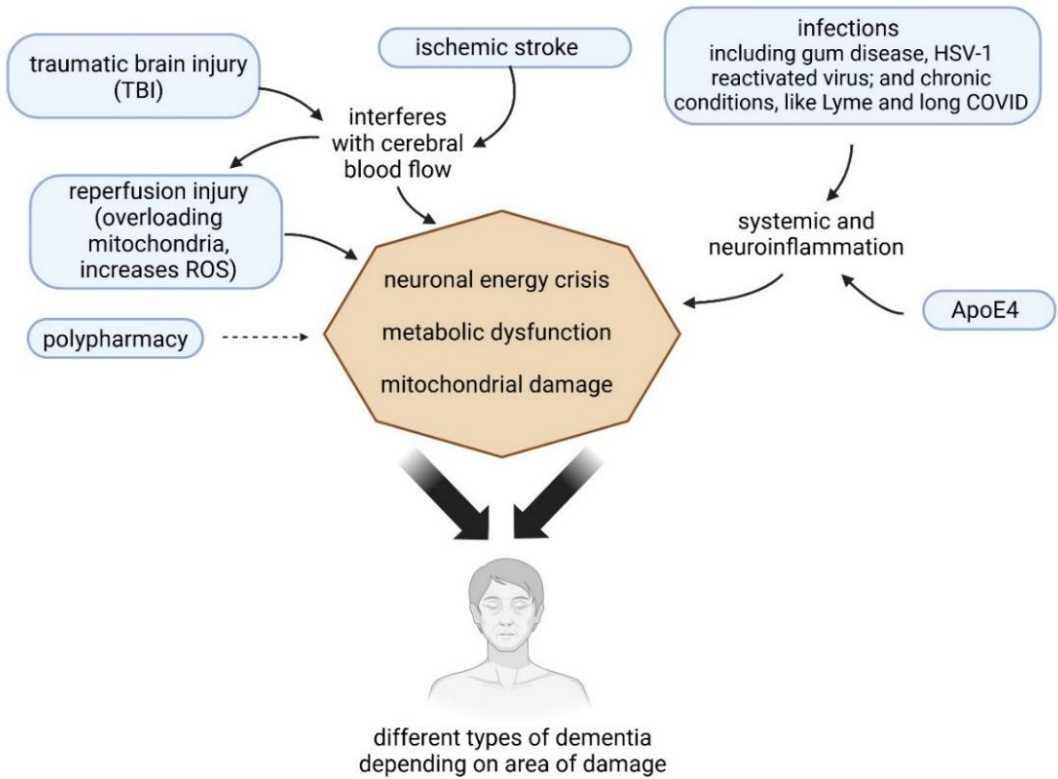

**Figure 3**. Framework of multi-factorial dementia emergence with neuronal energy crisis model as a mechanistic center point.

*Treatment and prevention strategies*

Fortunately, there do exist intervention strategies for some of these factors. For polypharmacy, these effects can be mitigated and often reversed through a better understanding of drug-drug interactions and through deprescription interventions of carefully reducing or stopping unnecessary medications.

For gum disease, a study by Cho and colleagues (134) found that preventive periodontal treatment may decrease mortality risk in older people with dementia, suggesting potential benefits of gum disease treatment on cognitive outcomes. Other anti-inflammatory interventions may also be of value, such as a recently reported study of 11745 dementia-free participants from the prospective population-based Rotterdam Study, where over a 14.5 year follow up, long term use of non-steroid anti-inflammatory (NSAID) medications was associated with a 12% reduction in the risk of developing dementia (135). Importantly, it was the long-term rather than intermittent use of NSAIDs that was associated with this benefit, suggesting that it is addressing long-term chronic inflammation that may be one of the pieces of the puzzle of mitigating dementia onset.

Antidiabetic treatments have also shown promise in reducing dementia risk (136). For instance, a large population-based cohort study found that initiation of SGLT-2 inhibitors was associated with a 35% lower risk of dementia compared with DPP-4 inhibitors in people with type 2 diabetes aged 40-69 years (137). Treatment with GLP-1 receptor agonists has been associated with a reduced incidence of dementia in patients with type 2 diabetes (138,139).

Dietary strategies have been extensively studied for their potential to mitigate cognitive decline and reduce the risk of neurodegenerative diseases like Alzheimer's. Caloric restriction, which involves reducing daily caloric intake without malnutrition, has been associated with improved metabolic profiles and reduced systemic inflammation. These effects may confer neuroprotection by enhancing mitochondrial function, reducing oxidative stress, and promoting autophagy, a cellular cleanup process that removes damaged proteins and organelles. Animal studies have demonstrated that CR can improve synaptic plasticity and cognitive function (140), while human studies suggest potential benefits in metabolic health markers linked to brain aging (141,142). Dr. Richard Isaacson, a prominent neurologist specializing in Alzheimer's prevention, has advocated for personalized nutritional interventions (143). These strategies include incorporating nutrient-dense foods, managing insulin resistance, and addressing individual metabolic needs.

Intriguingly, intranasal insulin delivery has been investigated for cognitive enhancement. Despite the fact that GLUT1 and GLUT3 transporters on BBB and neurons do not require insulin, it is needed for function of astrocytes that support neural function (144). Intranasal administration allows insulin to bypass the blood-brain barrier, directly reaching the central nervous system. Once in the brain, insulin can modulate neuronal glucose uptake, enhance synaptic plasticity, and reduce neuroinflammation. Several studies have reported that intranasal insulin administration can lead to improvements in memory and cognitive performance in patients with Alzheimer's disease and mild cognitive impairment. For instance, a randomized controlled trial demonstrated that a 4-month treatment with intranasal insulin improved delayed memory and functional abilities in adults with amnestic mild cognitive impairment or mild Alzheimer's disease (145).

Finally, photobiomodulation (PBM) therapy may come the closest to targeting the direct proposed culprit of mitochondrial dysfunction: the cytochrome c oxidase enzyme. It has gained attention for its potential to enhance mitochondrial function and provide neuroprotection (146,147). In PBM, a near-infrared light (NIR, ~810 nm) penetrates the skull and activates cytochrome c oxidase (Complex IV in the electron transport chain). This boosts ATP production and reduces oxidative stress in neurons. Preliminary studies suggest that PBM can improve cognitive function in patients with dementia (148). However, more extensive clinical trials are needed to establish the efficacy and optimal parameters of PBM therapy for neurodegenerative diseases.

Hearing and vision loss are linked to increased dementia risk due to reduced sensory stimulation, since sensory deprivation reduces cortical stimulation. Hearing loss increases cognitive load, and brain regions must compensate, increasing metabolic stress. However, mitigating these deficits through, for instance, hearing aids, has been shown to improve the situation (149,150). And finally, even such simple interventions as increased movement, like going for walks or dancing, have been shown to be protective against dementia.

Dementia is often seen as inevitable, but modern advancements allow for early detection of subtle cognitive and metabolic changes. Early detection enables intervention before reaching a critical threshold of irreversible neurodegeneration. Key areas of intervention include glucose regulation (preventing metabolic crisis in neurons); maintaining sensory input (vision, hearing); physical activity & cardiovascular health (supporting brain perfusion); as well as emerging pharmacological approaches, which target blood flow and mitochondrial function.

Modern technologies are enabling exciting tools for early detection. For instance, subtle motor changes often precede cognitive symptoms by years. Some dementia-related movement dysfunctions include reduced gait speed (slower walking); gait variability (i.e., Inconsistent step length and timing); dual-task performance decline (i.e., difficulty walking while performing a cognitive task), as well as postural instability (151,152). This occurs because dementia affects multiple brain regions that control movement, including the basal ganglia & frontal cortex, which are involved in motor planning and executive function; cerebellum, which controls balance and coordination, as well as white matter deterioration, leading to slower processing speeds affecting motor output. Neurodegeneration in these regions results in gait irregularities before memory impairment becomes noticeable. Interestingly, wearable sensors & AI-powered gait analysis can track these changes and predict dementia risk (153–155).

Pharmacologically, one exciting prospective treatment for Alzheimer's is exogenous administration of klotho, an endogenously produced protein that has emerged as an exciting and potentially transformative therapeutic candidate in the context of aging and neurodegenerative diseases, including Alzheimer's disease (AD). First discovered serendipitously in 1997 by Japanese scientist Makoto Kuro-o (156), the gene was named after Clotho, the Greek Fate who spins the thread of life, owing to its profound effects on longevity. Mice lacking klotho exhibited a syndrome resembling premature aging, while mice overexpressing klotho lived significantly longer—up to 30% more—highlighting its role as a longevity factor (157).

Klotho levels decline naturally with age in both mice and humans (158,159). This decline may contribute to cognitive deterioration and increased vulnerability to neurodegeneration. Mechanistically, klotho is primarily produced in the kidney and brain (notably in the choroid plexus), and it circulates in a soluble form with hormone-like actions (160). Factors such as chronic stress and epigenetic modifications (e.g., hypermethylation of the klotho gene promoter) have been associated with reduced expression (161,162).

Preclinical studies have shown that exogenous administration of soluble klotho can enhance cognition in both mice and nonhuman primates (163,164). Remarkably, a single injection of klotho improved memory performance in aged rhesus macaques, with effects lasting for 2–3 weeks—despite klotho's inability to cross the blood-brain barrier directly (164). This suggests that peripheral klotho may act through intermediate signaling molecules. One proposed pathway involves platelet factor 4 (PF4): klotho induces the release of PF4 from platelets, which can cross into the brain and enhance synaptic plasticity via GluN2B-containing NMDA receptors (165). However, even in PF4-knockout mice, klotho retains its cognitive benefits, suggesting that PF4 is sufficient but not necessary for its effects. Other mediators are currently under investigation.

These findings underscore klotho's broad neuroprotective and cognitive-enhancing potential, possibly acting like a "helmet" that shields neurons from age- and disease-related insults. Although human clinical trials are still needed, the evidence from rodent and primate studies supports further exploration of klotho as a multi-target therapy—one that enhances neural resilience rather than targeting a single pathology.

A summary of the key of the aforementioned therapeutic approaches is given in Table 1.

**Table 1**. Summary of emerging therapeutic approaches.

| Therapy | Target Mechanism | Clinical Impact |
| --- | --- | --- |
| **Intranasal Insulin** | Astrocytic glucose support | Memory improvement in patients with mild cognitive impairment (MCI) |
| **Klotho protein** | Systemic resilience, NMDA receptor tuning | Cognitive boost in non-human primates with cognitive deficits |
| **SGLT-2 & GLP-1 drugs** | Glucose regulation, anti-inflammatory signaling | Reduced dementia risk, ongoing trials |
| **NSAIDs (long-term)** | COX-2 inhibition, cytokine reduction | Lower AD incidence in some populations |
| **Photobiomodulation (PBM)** | Mitochondrial activation via red/NIR light | Increased energy metabolism, reduced Aβ |

Dementia may be hard to reverse at later stages, when a lot of the damage has already taken place. However, with early enough detection and a better understanding of which pathway may be leading to it (metabolic syndrome, vascular disease, trauma, infection, etc.), it may be possible to if not prevent it, then at least mitigate the time of onset. The mathematical framework presented here may be useful to better understand the relative weight of different pathways to pathology, as well as different interventions for an individual's emerging pathology, complementing other tools to delay symptoms and extend patient's quality of life.

**Conflicts of interest**

IK is an employee of EMD serono, a US business of Merck KGaA. The views presented in this paper are the author's personal views and do not necessarily represent the views of EMD Serono. This research received no financial support.